\documentclass[preprint,aps,showpacs]{revtex4}
\begin{document}
\preprint{UCI-TR 2003-41}
\title{Expanding Cosmologies in Brane Geometries}
\author{Myron Bander\footnote{Electronic address: mbander@uci.edu}
}
\affiliation{
Department of Physics and Astronomy, University of California, Irvine,
California 92697-4575}

\date{August\ \ \ 2003}

\begin{abstract} Five dimensional gravity coupled, both in the bulk
and on a brane, to a scalar Liouville field yields a geometry confined
to a strip around the brane and with time dependent scale factors for
the four geometry. In various limits known models can be recovered as
well as a temporally expanding four geometry with a warp factor
falling exponentially away from the brane. The effective theory on the
brane has a time dependent Planck mass and ``cosmological
constant''. Although the scale factor expands, the expansion is not
an acceleration.
\end{abstract}
\maketitle
There is considerable interest in theories of gravity and of
cosmology with extra dimensions where our world is confined to a four
dimensional space-time subspace or 3-brane. All of our known fields,
with the exception of gravity, are confined to the brane. The extra
dimensions may have compact toroidal topologies
\cite{Arkani-Hamed:1998rs} or be unbounded with a scale factor, warp, 
depending on the ``distance'' from the brane \cite{Randall:1999ee,
Randall:1999vf}. Branes meandering in the internal space, with the
brane metric dependent on the internal coordinates, were treated in
Ref. \cite{Kaloper:1999sm}. Less singular metrics we obtained in
situations where the branes were thickened \cite{Csaki:2000fc}. As
many of these works concerned themselves with the hierarchy problem they
restricted themselves to Minkowski metrics on the brane; specifically,
the metrics were time independent. Extending these concepts to
cosmology requires the introduction of time dependent scale factors on
the brane and possibly in the extra dimensions. Solutions in which one
posits various stress tensors on the brane and a general review of
cosmology restricted to a brane in extra dimensions may be found in
Ref. \cite{Langlois:2002bb}.

In this work we look at a five dimensional bulk whose dynamics is
governed by a scalar Liouville field coupled to gravity in the usual
way.  In addition there is a coupling to the scalar field and to the 
tension on a thin 3-brane. As in previous works the brane tension is
finely tuned to parameters of the bulk action. The general form of the
metric we obtain is
\begin{equation}\label{metric1}
ds^2=\left(1-\frac{|y|}{y_0}\right)^\xi
  \left[dt^2-a(t)^2d{\bf x}^2\right]
   -b(t)^2dy^2\, ,
\end{equation} 

In the extra dimension the bulk geometry is confined to a finite
strip, $y\le y_0$, around the brane of interest. Although $y_0$ may be
scaled away (set equal to one), we keep it for the convenience of
limiting procedures discussed further on. It will turn out that for
$\xi\ge 1/2$ we may ignore singularities at the edges of the strip;
for $\xi<1/2$ these singularities force us either to identify the
opposite edges of the strip and place a regulator brane at $y=y_0$ or
place two branes at $y=\pm y_0$ . As the metric vanishes on these
extra branes they do not support any physics.

In addition to the trivial solution with $a(t)$ and $b(t)$ in
(\ref{metric1}) being constant in time, the {\it ansatz}
$a(t)=a_0(t/t_0)^\alpha$, $b(t)=b_0(t/t_0)^\beta$ yields a solution
provided
\begin{equation}\label{ansatz}
\alpha=\frac{2+3\xi\pm 2\sqrt{1+\xi}}{8+9\xi}\, , \quad 
   \beta=\frac{6\mp 6\sqrt{1+\xi}}{8+9\xi}\, .
\end{equation}
These satisfy the  relation $\beta=1-3\alpha$, reminiscent of one of the
Kasner \cite{Kasner} conditions. For the upper solution, $\alpha$
ranges from 1/2 to 1/3 and $\beta$ from -1/2 to 0 as $\xi$ goes from 0
to infinity. for the lower solution $\alpha$ goes from 0 to 1/3 and
$\beta$ from 1 to 0.

There are various interesting limits. In addition to being able to
recover the geometry of \cite{Randall:1999ee,Randall:1999vf} we can
obtain a cosmology where the four metric represents an expanding
universe with a warp factor decreasing exponentially as we move away
from the brane 
\begin{equation}\label{limitmetric}
ds^2=e^{-2k|y|}\left[dt^2-a_0\left(\frac{t}{t_0}\right)^
{\frac{2}{3}}d{\bf x}^2\right]-dy^2\, .
\end{equation}
This limit is interesting as we recover an effective four dimensional
cosmology with a time independent Planck mass. For $\xi=0$ we recover
the Kasner solutions. We shall return to a discussion of these metrics
further on.

The solution (\ref{metric1}) is obtained from the action for the metric
tensor and for a scalar Liouville field with contributions from the
bulk and from one or two branes. Five dimensional theories with bulk
scalar fields have been previously considered. A massive scalar field
can determine the size of the internal dimension
\cite{Goldberger:1999wh} and with intricate self couplings can thicken 
the branes \cite{Gremm:1999pj, Bazeia:2003aw}.

The contribution from one of the branes, presumably the one we are on,
will be indicated explicitly while the one from the other brane or
branes will be left for later elaboration.
\begin{eqnarray}\label{S}
S&=&S_{\rm bulk}+S_{\rm brane}\, ,\nonumber\\
S_{\rm bulk}&=&\frac{1}{2k_5^2}\int d^5x\, \sqrt{g}\left\{
  R+\frac{1}{2}\partial_\mu\phi\partial_\nu\phi g^{\mu\nu}+
  \lambda\exp({-\kappa\phi})\right\}\, ,\\
S_{\rm brane}&=&\frac{h}{2k_5^2}\int d^5x\, \sqrt{g}\, 
\delta(n_\mu x^\mu)\sqrt{n_\mu n_\nu g^{\mu\nu}}
\exp({-\frac{\kappa}{2}\phi})\, ;\nonumber
\end{eqnarray}
$k_5=8\pi/M_5^3$ where $M_5$ is the five dimensional Planck
mass. $\kappa$ is a free parameter and although $\lambda$ is included
for convenience it can be scaled by any positive number through a
shift in the field $\phi$. In $S_{\rm brane}$ $n_\mu$ is a spacelike
vector normal to the brane and the product $\delta(n_\mu
x^\mu)\sqrt{n_\mu n_\nu g^{\mu\nu}}$ is independent of the magnitude
of this vector. Varying the combination $\sqrt{g}\, \sqrt{n_\mu n_\nu
g^{\mu\nu}}$ with respect to $g^{\mu\nu}$ yields terms proportional to
$(-g_{\mu\nu}+ n_\mu n_\nu/n_\alpha n_\beta g^{\alpha\beta})$, namely
depending only on the metric along the brane; this procedure leads to
the same results as one would get by using the Israel junction
conditions \cite{Israel}. As in all previous works we will take
$n_\mu$ to be along $x_5$ for which we will use the symbol $y$.  Note
that the $\phi$ coupling in $S_{\rm brane}$ is $\kappa/2$ as opposed
to $\kappa$ in $S_{\rm bulk}$. The magnitude of the brane tension,
determined by $h$, as in previous works, is related to bulk parameters;
our solutions require
\begin{equation}\label{h}
h=\sqrt{\frac{2\lambda}{\kappa^2\left(8/3\kappa^2-1
  \right)}}\, ;
\end{equation} this restricts $\lambda\ge 0$ for $\kappa^2\le8/3$ and
$\lambda<0$ otherwise.

The solution for the equations of motion obtained by varying (\ref{S})
with respect to $g^{\mu\nu}$ and $\phi$ we seek have a metric given in
({\ref{metric1}) and the scalar field of the form
\begin{equation}\label{scalar}
\phi=A\ln\left[\left(1-\frac{|y|}{y_0}\right)b(t)\right]-C\, .
\end{equation}
It is straightforward to check that for $y_0>|y|$ these are indeed
solutions provided
\begin{eqnarray}\label{cond1}
A&=&2/\kappa\, ,\nonumber\\
C&=&\frac{1}{\kappa}\ln\left[\frac{2}{\lambda\kappa^2y_0^2}
   \left(\frac{8}{3\kappa^2}-1\right)\right]  \, ,\\
\xi&=&\frac{4}{3\kappa^2}\nonumber
\end{eqnarray}
and the scale factors $a(t)$ and $b(t)$ satisfy
\begin{eqnarray}\label{cond2}
4\left(\frac{\dot a}{a}\right)^2+
  4\left(\frac{\dot a}{a}\right)\left(\frac{\dot b}{b}\right)
    -\xi\left(\frac{\dot b}{b}\right)^2&=&0\, ,\nonumber\\
4\frac{\ddot a}{a}+4\left(\frac{\dot a}{a}\right)^2
    +\xi \left(\frac{\dot b}{b}\right)^2&=&0\, ,\\
8\left(\frac{\ddot a}{a}\right)^2+4\left(\frac{\ddot b}{b}\right)^2
   +4\left(\frac{\dot a}{a}\right)^2+
     8\left(\frac{\dot a}{a}\right)\left(\frac{\dot b}{b}\right)
      +3\xi \left(\frac{\dot b}{b}\right)^2&=&0\, ;\nonumber
\end{eqnarray} the dot represents differentiation with respect to
time. 

With the choice of metric in (\ref{metric1}), of the twenty five
equations for the components of the Einstein tensor,
$R_{\mu\nu}-\cdots=0$ and the equation of motion for the field $\phi$
only only five are nontrivial and independent. These may be chosen to
be the equation of motion for $\phi$ and for the $tt$, $ty$, $yy$ and
any of the diagonal space-space component of the Einstein tensor along
the brane. The relations between $A$, $\kappa$ and $\xi$ in
(\ref{cond1}) solve the $ty$ equation while (\ref{cond2}) takes care
of the other four. That, in the bulk, these {\em four} equations yield
only the {\em three} conditions in (\ref{cond2}) is not surprising as
the equation of motion for the field and the Einstein equations for
the metric are related by the conservation of the energy-momentum
tensor. What is pleasant is that all the four independent equations on
the brane, the ones involving $\delta(y)$ terms, where the
energy-momentum tensor is not conserved, are also satisfied.

We now turn to possible singularities at $|y|=y_0$. For $\xi> 1/2$ or
equivalently $\kappa < \sqrt{3/8}$ we can restrict the bulk to the
strip $|y|\le y_0$ as the solutions discussed above may be continued to the
end points. For $\xi\le 1/2$ singularities develop at these points and
the solutions are no longer valid there. As in many previous
discussions of bulk-brane geometries the cure consists of either
identifying $y=y_0$ with $y=-y_0$ (orbifolding) and introducing a
brane at $|y|=y_0$ or introducing independent branes at $y=\pm
y_0$. In the first case, the action on the $y=y_0$ brane is,
\begin{equation}\label{brane'}
{S'}_{\rm brane}=-\frac{h}{2k_5^2}\int d^5x\, \sqrt{g}\, 
\delta(n_\mu x^\mu-y_0)\sqrt{n_\mu n_\nu g^{\mu\nu}}
\exp({-\frac{\kappa}{2}\phi})\, .
\end{equation}
If one wishes to place branes at both ends of the strip, the action
contributed is one half that of (\ref{brane'}) on each of the two
branes.  Since the four-metric in (\ref{metric1}) vanishes at $|y|=y_0$,
these branes or brane cannot support any physics. The explicit forms
for $a(t)$ and $b(t)$ are given in (\ref{ansatz}). For $\xi\ge 2$ the
edge $|y|=y_0$ is at the horizon in that it takes an infinite time to
reach it.

Certain limits of these solutions are interesting. The case $\xi=0$
corresponds to Kasner's \cite{Kasner} solutions. For the $a(t)$ and
$b(t)$ constant case the limit $\xi\rightarrow\infty$ with
$y_0=\xi/(2k)$ yields the Randall-Sundrum solution
\cite{Randall:1999ee, Randall:1999vf}. Eq. (\ref{h}) is equivalent to
their relation between the bulk cosmological constant and brane
tension. In the same limit, but with $\alpha$ and $\beta$ given by
either solution in (\ref{ansatz}) we obtain the metric
(\ref{limitmetric}) and
\begin{equation}
\phi(t,y)=2\frac{\ln(t)}{\sqrt{3}}\, .
\end{equation}
The above solution may be obtained independently from the action
$S=S_{\rm bulk} +S_{\rm brane}$, where
\begin{eqnarray}\label{expact}
S_{\rm bulk}&=&\frac{1}{2k_5^2}\int d^5x\, \sqrt{g}\left\{
  R+\frac{1}{2}\partial_\mu\phi\partial_\nu\phi g^{\mu\nu}+
  3k^2\right\}\, ,\nonumber\\
S_{\rm brane}&=&\frac{3k}{4k_5^2}\int d^5x\, \sqrt{g}\, 
\delta(n_\mu x^\mu)\sqrt{n_\mu n_\nu g^{\mu\nu}}\, .
\end{eqnarray}

How well gravity on the brane is described by $ds^2=dt^2-a(t)^2d{\bf
x}^2$ depends on solution of the equation
\begin{equation}\label{fluctuation}
-\left(1-\frac{|y|}{y_0}\right)^{-2\xi}\partial_y\left[
  \left(1-\frac{|y|}{y_0}\right)^{2\xi}\partial_y\right]h(y,t)=
   m^2b(t)^2h(y,t)\, ,
\end{equation} 
with $h(y,t)$ describing fluctuations around the metric. Fluctuation
equations for non zero $m$ are quite complex (see for
example Refs. \cite{DeWolfe:1999cp, Giddings:2000mu}) and here we shall
restrict our study to $m=0$; it is necessary to have an acceptable
solution for this case and it is easy to exhibit such a solution
\begin{equation}
h_{m=0}(y,t)\sim\epsilon(y)\left[
\left(1-\frac{|y|}{y_0}\right)^{1-2\xi}-1\right]\, ;
\end{equation}

Four dimensional gravity on the brane appears after integrating
the action (\ref{S}) with the metric
\begin{equation}\label{4metric}
ds^2=\left(1-\frac{|y|}{y_0}\right)^\xi {}^{(4)}g_{ij}(x)dx^idx^j-b(t)^2dy^2\, 
\end{equation}
over $y$; ${}^{(4)}g_{ij}(x)$ is the four metric on the brane. This may be
accomplished by using the ADM reduction with $b(t)$ playing the role
of the lapse function and conformally transforming the resulting four
dimensional metric by the factor
$\left(1-\frac{|y|}{y_0}\right)^{-\xi}$. The result is
\begin{eqnarray}\label{4action}
S&=&\frac{1}{2k_5^2}\int d^4xdy\sqrt{-{}^{(4)}g}
  \left(1-\frac{|y|}{y_0}\right)^{2\xi}b(t)\left[
   \frac{R\left({}^{(4)}g_{ij}(x)\right)}{\left(1-\frac{|y|}{y_0}\right)^\xi}
   +\frac{3\xi}{2\left(1-\frac{|y|}{y_0}\right)^\xi}\left(\frac{\dot
   b}{b}\right)^2
     \right]\, ,\nonumber\\
&=&\frac{1}{2k_5^2}\int d^4x\frac{2y_0b(t)}{\xi+1}\sqrt{-{}^{(4)}g}
\left[R\left({}^{(4)}g_{ij}(x)\right)+\frac{3\xi}{2}\left(\frac{\dot
b}{b}\right)^2
  \right]\, .
\end{eqnarray}
The four dimensional theory has a time dependent Planck mass
\begin{equation}\label{4planckmass}
M_4(t)^2=M_5^3\frac{2y_0b(t)}{\xi+1}
\end{equation}
and a time dependent ``cosmological constant''
\begin{equation}\label{4const} 
\frac{\Lambda(t)}{M_4(t)^2}\sim
\xi\beta^2\left(\frac{t}{t_0}\right)^{\beta -1}\, .  
\end{equation}
In cosmologies small extra dimensions, when known physics is not
restricted to a brane,time dependence of the internal dimensions is
severely restricted by limits on the temporal variations of all
fundamental constants \cite{Kolb:1985sj}; in contrast for theories
with most of known physics restricted to a brane, only limits on the
time evolution of the Planck mass may come into play. The solutions
discussed here can accommodate any such limits as by choosing $\kappa$
small and equivalently $\beta$ small we can make this variation as
soft as necessary. Temporal variations of the cosmological constant,
or more generally of the dark energy are coming into
consideration\cite{Page:2003pn}.

The solutions presented have $\alpha\le 1/2$ and thus represent
decelerating cosmologies. In line with recent observations
\cite{Page:2003pn} we would like to accommodate an accelerating, 
expanding scale factor. Having a time dependent scale factor for the
external dimensions circumvents some no-go theorems
\cite{Townsend:2003fx, Wohlfarth:2003kw}. In thee present context we 
can achieve accelerating scale factors by analytically continuing the
solutions to negative $\xi$. This, however corresponds to an imaginary
exponent in the Liouville action. Whether this difficulty can be
circumvented is under investigation. Difficulty in finding
accelerating solution was noted in Ref. \cite{Wohlfarth:2003kw}.

\end{document}